\documentclass[10pt,letterpaper]{article}
\usepackage[T1]{fontenc}
\usepackage[top=0.85in,left=1.5in,footskip=0.75in,marginparwidth=1.5in]{geometry}
\usepackage[justification=justified]{caption}

\usepackage{xcolor}
\usepackage{svg}
\usepackage{ragged2e}
\usepackage{amsfonts}
\usepackage{amsmath}
\usepackage{float}

\usepackage{tabularx}
\usepackage{array}
\usepackage{booktabs}

\usepackage{hyperref}

\usepackage[utf8]{inputenc}

\usepackage{cite}
\usepackage{booktabs}
\usepackage{tabularx}
\usepackage{array}
\usepackage{ragged2e}

\newcolumntype{Y}{>{\RaggedRight\arraybackslash}X}
\usepackage{nameref,hyperref}

\usepackage{xcolor}

\usepackage{soul}
\usepackage{xcolor}

\raggedright
\usepackage{setspace} 
\usepackage{changepage}

\usepackage[aboveskip=1pt,labelfont=bf,labelsep=period,singlelinecheck=off]{caption}

\makeatletter
\renewcommand{\@biblabel}[1]{\quad#1.}
\makeatother

\usepackage{lastpage,fancyhdr,graphicx}
\usepackage{epstopdf}
\usepackage{xcolor}
\usepackage{float}

\definecolor{Gray}{gray}{.25}

\usepackage{xcolor}
\definecolor{myorange}{RGB}{230,120,0}

\usepackage{graphicx}

\usepackage{sidecap}

\usepackage{wrapfig}
\usepackage[pscoord]{eso-pic}
\usepackage[fulladjust]{marginnote}
\reversemarginpar

\usepackage{array}
\usepackage{caption}
\usepackage{booktabs}
\usepackage{longtable}
\usepackage{array}
\usepackage{ragged2e}
\usepackage{pdflscape}
\usepackage{caption}

\newcommand{\Hsys}{H}
\newcommand{\Asys}{A}
\newcommand{\HA}{H{+}A}
\newcommand{\Hpre}{H_{\mathrm{pre}}}
\newcommand{\Hpost}{H_{\mathrm{post}}}
\newcommand{\Coupling}{C}
\newcommand{\System}{\mathcal{S}_{H\text{-}AI}}

\begin{document}
\vspace*{0.2in}
\justifying

% title goes here:
\begin{flushleft}

{\Large Human-LLM Cognitive Extension: A Framework for Studying Coupled Cognition}

% authors go here:
\bigskip
\bigskip

{\large Ernesto Biempica$^{1*}$,
Emiliano Bruner$^{2,3\dagger}$,
Juan F. Poyatos$^{2,4\ddagger}$}
\\

$^1$ Universitat Autònoma de Barcelona, Barcelona 08193, Spain\\
$^2$ Museo Nacional de Ciencias Naturales (MNCN-CSIC), Madrid 28006, Spain\\
$^3$ Reina Sofia Foundation Alzheimer Centre, CIEN Foundation, ISCIII, Madrid, Spain\\
$^4$ Vienna Centre for Advanced Studies (ViCAS), Vienna 1010, Austria\\
$^*$\texttt{ernestojose.biempica@autonoma.cat}\\
$^\dagger$ \texttt{emiliano.bruner@mncn.csic.es}\\
$^\ddagger$ \texttt{juanfpoyatos@csic.es}\\

\end{flushleft}

\vspace{.25cm}
\noindent
{\large\bfseries Abstract}
\vspace{.15cm}

\noindent
Recent research on the use of Large Language Models (LLMs) reports divergent cognitive outcomes, including improved assisted performance, limited transfer, and compromised unaided learning. Here, we argue that part of this seeming inconsistency arises because studies often assess different cognitive systems without making the unit of analysis explicit. Building on work in extended and distributed cognition, we propose the Human-LLM Cognitive Extension (HLCE) as a methodological framework for matching experimental designs, measures, and causal inferences to four distinct systems: the human acting independently, the artificial agent, the joined human-LLM configuration, and the human after division. The framework also treats the coupling architecture as part of the experimental object. HLCE separates performance augmentation, system-level complementarity, post-interaction human change, and stronger claims of constitutive cognitive integration. This separation can improve the interpretation of existing evidence and guide the design of human-LLM interactions without assuming in advance that these configurations constitute extended minds. We illustrate the application of HLCE with a synthetic diagnostic experiment comparing two connecting configurations that differ in how supporting information is communicated. The example highlights how distinct couplings  can shape the organization of joint problem-solving, even when the final outcome remains unchanged. Overall, HLCE provides a methodological basis for designing controlled experiments and interpreting evidence across distinct forms of human-LLM cognitive interaction.

\vspace{.25cm}


\newpage

%\vspace{.25cm}
%\noindent
%{\large {\bf Significance statement}}
%\vspace{.15cm}

%\noindent
%If required.
 
\newpage

%%%%%%%%%%%%%%%%%%%%%%
\section*{Introduction}
%%%%%%%%%%%%%%%%%%%%%%

Artificial intelligence (AI) is increasingly embedded in activities that depend on reasoning, learning, problem solving, and decision making. Large language models (LLMs) are particularly relevant in this context because they interact with users through language and participate directly in activities traditionally considered expressions of human cognitive performance~\cite{hernandez-orallo_enhancement_2025,smart_chatgpt_2025}. Yet evidence concerning their cognitive consequences is increasingly difficult to interpret in a unified way.

Indeed, LLM assistance can improve productivity and task performance while it is available~\cite{noy_experimental_2023,dellacqua_navigating_2026}, but these benefits do not necessarily transfer to subsequent unaided performance~\cite{wu_human-generative_2025}. In some settings, improved AI-assisted task performance has even been associated with poorer subsequent learning outcomes~\cite{stromberg_generative_2026}. Other studies show that the effects of AI assistance depend strongly on task characteristics and the interplay between human and artificial capabilities~\cite{dellacqua_navigating_2026,vaccaro_when_2024}. Taken together, these findings raise a fundamental question: what exactly is being improved, preserved, or degraded when the impact of AI use is assessed?

We argue that part of the apparent inconsistency arises because Human-AI studies do {\it not} always evaluate the same cognitive system. At least four analytically contrasting targets should be distinguished: the human acting independently ($H$), the artificial system acting independently ($A$), the coupled Human-AI configuration ($H+A$), and the human {\it after} AI assistance has been removed ($H_{\rm post}$). This distinction matters because improvement in one category does not imply improvement in another \cite{wu_human-generative_2025,vaccaro_when_2024}. For example, superior performance of $H+A$ relative to $H$ indicates an advantage of the assisted configuration but does not establish an improvement in the human's autonomous capability.

Conversely, a decline in $H_{\rm post}$ can coexist with enhanced capabilities while $H+A$ remains {\it coupled}. Statements such as ``AI improves cognition´´ or ``AI degrades cognition´´ are therefore ambiguous unless the system to which the observed effect belongs is made explicit. Existing Human-AI research already involves these contrasting targets, but they are not consistently recognized when findings are interpreted across studies~\cite{xu_applying_2024,zhao_understanding_2026,davis_interaction-centered_2026,stromberg_generative_2026}.

%\textcolor{myorange}{
%Culturally transmitted representational tools, such as mathematical notation, graphs, and geometric diagrams, may provide an illustrative case of persistent cognitive transformation. Sablé-Meyer et al. \cite{sable-meyer_geometric_2025} show that human geometric perception relies on compositional representations whose complexity can be characterized by minimum description length, suggesting that abstract structures can be internally encoded and manipulated. Within the Human--Tool Cognitive Extension framework, repeated exposure to external symbolic representations may similarly promote the internalization of new concepts and reasoning strategies that remain accessible after tool withdrawal. Such persistence would be reflected in $H_{\mathrm{post}}$, distinguishing temporary tool-assisted performance from lasting cognitive change. Although not directly demonstrated by Sablé-Meyer et al., this possibility offers a testable hypothesis concerning the enduring effects of human--tool interaction.}

Identifying the system under study is necessary but not sufficient, because $H+A$ is not a unique experimental condition. The properties of the Human-AI configuration also depend on how its components are combined. The same human, the same LLM, and the same task may produce contrasting outcomes depending on features such as information flow, initiative, transparency, reliability, persistence, personalization, or the available opportunities for reciprocal feedback~\cite{heersmink_dimensions_2015}. We therefore represent the {\it coupled system} at a given time as a function of the relevant states of the human and LLM components and their evolving interaction condition, which is also contingent on the architecture of the coupling, $\Coupling$. In this formulation, $\Coupling$ is not merely background context but part of the experimental object itself. Consequently, claims about Human-AI cognition should specify both which system is evaluated and under which configuration that system operates~\cite{xu_applying_2024,zhao_understanding_2026}.

This methodological distinction leads to a broader theoretical question. Demonstrating that an AI system affects human performance, or that a Human-AI arrangement works better than either component alone, does not by itself establish that the artificial component forms part of the cognitive mechanism responsible for that performance. The Extended Mind (EM) hypothesis provides a principled way to formulate this boundary problem~\cite{clark_extended_1998}. From this perspective, AI may serve as an external cognitive resource~\cite{clark_extending_2025}, while cognitive integration approaches emphasize that internal and external resources may contribute complementary functions to an integrated cognitive system~\cite{menary_cognitive_2010}. The relevance of these ideas to LLMs has been explored by Smart, Clowes and Clark~\cite{smart_chatgpt_2025}. Their ``Digital Andy'' system couples GPT-3.5 with an external vector database containing Clark's philosophical works through retrieval-augmented generation (RAG), drawing an analogy with the external memory illustrated by Otto's notebook in Clark and Chalmers~\cite{clark_extended_1998}.

%This methodological distinction leads to a broader theoretical question. Demonstrating that an AI system affects human performance, or that a Human-AI arrangement works better than either component alone, does not by itself establish that the artificial component forms part of the cognitive mechanism responsible for that performance. The Extended Mind (EM) hypothesis provides a principled way to formulate this boundary problem \cite{clark_extended_1998}. \textcolor{blue}{Indeed, AI can be easily acknowledged as peripheral extra-somatic device of the human mind \cite{clark_extending_2025}. Its inductive and intuitive mechanisms based on neural network, machine learning, multidimensional spaces, and the principle of correlation, can be well matched with the deductive and conceptual nature of the human reasoning \cite{bruner_self_2026}.}
%\textcolor{myorange}{Smart, Clowes and Clark~\cite{smart_chatgpt_2025} provided one of the first explicit attempts to reconnect the classical EM framework with contemporary LLMs through an implemented system. Their “Digital Andy” system couples GPT-3.5 to an external vector database containing Clark's philosophical works through retrieval-augmented generation (RAG), drawing an analogy between the LLM memory and the external memory exemplified by Otto's notebook in Clark and Chalmers \cite{clark_extended_1998}. }

Rather than assuming that Human-AI interaction constitutes extended cognition, we use the EM concept to distinguish external causal support from {\it stronger} claims of functional and constitutive cognitive integration. Previous work on multidimensional cognitive integration and mechanistic criteria for demarcating cognitive boundaries further suggests that the degree and organization of coupling, rather than merely the presence of an external tool, should be subjected to empirical examination. The EM hypothesis therefore serves here not as a predefined characterization of Human-LLM systems, but as a framework for asking when an artificial component is merely useful for cognition and when it may participate in the mechanism that realizes it.

On this basis, we introduce the Human-LLM Cognitive Extension ({\tt HLCE}) framework as a methodological framework for studying Human-LLM cognition. {\tt HLCE} has five functions: to specify the cognitive system to which an observed outcome is attributed; to characterize and, where possible, experimentally manipulate the Human-LLM coupling architecture; to match cognitive measures to the appropriate level of analysis; to distinguish performance while coupled from changes that persist {\it after} decoupling; and, when constitutive cognitive integration is itself the hypothesis, to propose causal tests that distinguish reciprocal interaction from stronger evidence of mechanistic integration.

%\textcolor{myorange}{Beyond its methodological purpose, characterizing the conditions under which human–LLM cognitive integration occurs may also provide insights for the future design of AI systems that more effectively support human problem-solving and learning processes, particularly by promoting meaningful cognitive engagement rather than merely optimizing task performance.}

%{\tt HLCE} is therefore {\it not} intended to assume that Human-LLM systems constitute extended minds, nor to replace existing approaches to Human-AI collaboration investigations \cite{smart_chatgpt_2025,hernandez-orallo_enhancement_2025}. Its purpose is to provide a common methodological structure for determining what system has changed, under what form of coupling, and what conclusions the resulting evidence can support.

{\tt HLCE} is therefore {\it not} intended to assume that Human-LLM systems constitute extended minds, nor to replace existing approaches to Human-AI collaboration research \cite{smart_chatgpt_2025,hernandez-orallo_enhancement_2025}. Its purpose is to provide a common methodological structure for determining which system is being evaluated, how it is coupled, and what conclusions the resulting evidence can support. Beyond this methodological role, {\tt HLCE} may also inform the design of Human-LLM interactions that support human problem-solving and learning, by directing attention to cognitive engagement and the potential for lasting changes in human capabilities rather than assisted task performance alone.

%%%%%%%%%%%%%%%%%%%%%%%%%%%%%%%%%%%%%%%%%%%%%%%%%%%%%%
\section*{From interaction to cognitive integration}
%%%%%%%%%%%%%%%%%%%%%%%%%%%%%%%%%%%%%%%%%%%%%%%%%%%%%%

Considering a Human-AI arrangement as a testing object does not imply that the artificial component is part of the cognitive system. Extrinsic resources can alter function, reduce cognitive burden, or provide information while remaining causally {\it external} to the mechanism that implements the relevant cognitive process~\cite{clark_extended_1998,kaplan_how_2012}. The theoretical problem is therefore not whether an LLM influences cognition---it clearly can---but under what circumstances that effect warrants treating the artificial component as functionally or constitutively integrated into the cognitive system.

%\textcolor{myorange}{Kaplan \cite{kaplan_how_2012} addresses this distinction through the criterion of \textit{mutual manipulability}, proposing that a component's constitutive relevance to a cognitive mechanism can be investigated through reciprocal interventions on the component and the system-level cognitive activity. This provides a methodological basis for experimentally examining whether an LLM merely influences human cognition or participates in an integrated cognitive mechanism.}

%The EM notion provides one scheme for addressing this problem. Clark and Chalmers~\cite{clark_extended_1998} challenged the premise that the boundaries of cognition must be the same as those of the biological organism. On a {\it functionalist} account, external elements may participate in cognition when they become sufficiently integrated into the processes through which cognitive functions are achieved. The canonical example of Otto's notebook illustrates the possibility that information stored outside the brain can, under appropriate conditions, operate as part of an individual's memory system. Therefore, the implication for Human-AI research is not that every extrinsic support constitutes cognition, but rather that the boundaries of cognition should be regarded as a theoretical and empirical exploration not fixed in advance.

The EM notion provides one scheme for addressing this problem. Clark and Chalmers~\cite{clark_extended_1998} challenged the premise that the boundaries of cognition must be the same as those of the biological organism. On a {\it functionalist} account, external elements may participate in cognition when they become sufficiently integrated into the processes through which cognitive functions are achieved. The canonical example of Otto's notebook commented above illustrates the possibility that information stored outside the brain can, under appropriate conditions, operate as part of an individual's memory system. Therefore, the implication for Human-AI research is not that every extrinsic support constitutes cognition, but rather that the boundaries of cognition should be regarded as a theoretical and empirical question, not fixed in advance.

%\textcolor{myorange}{Wilson \cite{wilson_re-tooled_2010} further argues that culturally acquired cognitive tools can reorganize the mechanisms underlying cognitive activity, rather than merely providing external support. This distinction is particularly relevant to Human–LLM interaction, where changes in cognitive routines may provide evidence of cognitive reorganization, without necessarily demonstrating constitutive cognitive extension.}

This distinction also requires differentiating cognitive {\it offloading} from cognitive {\it extension}, without treating them as mutually exclusive categories. Cognitive offloading refers to the use of external resources to reduce the cognitive demands of a task, whereas cognitive extension concerns the conditions under which such resources become constitutive components of a cognitive mechanism. An external resource may improve performance by replacing or reducing internal cognitive operations without thereby becoming part of the mechanism performing them. Offloading may therefore occur within an extended cognitive system, but does not, by itself, establish constitutive integration. This distinction is especially important for LLMs because superior assisted performance may arise from delegation rather than integration. A Human-LLM combination may thus be more capable than the unaided human while providing no evidence that the human's cognitive capacities have developed or that the LLM has become part of an extended cognitive mechanism. Sustained tool use can nevertheless reorganize users' own cognitive strategies over time~\cite{wilson_re-tooled_2010}, making post-interaction change a separate empirical question rather than an implication of assisted performance.

Cognitive integration should also not be viewed as an all-or-none feature. Heersmink's multidimensional description characterizes human-artifact integration along dimensions including reciprocal information flow, reliability, durability, trust, transparency, individualization, and transformation~\cite{heersmink_dimensions_2015}. For {\tt HLCE}, the relevance of these dimensions is {\it methodological}: they provide candidate properties through which the coupling architecture $C$ can be described and, for some dimensions, experimentally manipulated. They should not themselves be regarded as confirmation that extension has occurred.

A stronger claim, i.e., that an artificial component participates constitutively in a cognitive mechanism, requires evidence beyond association, improved function, or bidirectional communication. Kaplan's mechanistic approach suggests {\it mutual manipulability} as a criterion for investigating cognitive boundaries: intervention on a possible component should systematically affect the phenomenon produced by the mechanism, while interventions directed at the mechanism should in turn systematically affect the candidate component~\cite{kaplan_how_2012}. In the case of Human-LLM interaction, this motivates intervention-based tests of reciprocal causal dependence, while cautioning against considering simple message effects or behavioral correlations as adequate evidence of constitutive integration. Even reciprocal intervention effects must be interpreted carefully, since causal dependence does not by itself establish constitutive participation.

Taken together, these positions separate increasingly stronger claims about Human-AI interaction. An artificial system may first provide {\it causal support}, enabling improved performance without becoming part of the cognitive mechanism. More tightly associated systems may exhibit {\it reciprocal coordination or dependence}. Stronger still are forms of {\it functional integration}, in which cognitive operations are systematically distributed across elements \cite{heersmink_dimensions_2015,kaplan_how_2012}. Finally, claims of {\it constitutive cognitive extension} require evidence that the external component participates in the mechanism realizing the cognitive process itself. These levels should not be inferred from one another automatically. {\tt HLCE} uses this distinction to align theoretical claims with the evidence required to support them.

A concise conceptual genealogy of the EM framework is provided in Table~\ref{table:S1}; the present analysis focuses on the experimental implications of system individuation and coupling architecture.

%%%%%%%%%%%%%%%%%%%%%%%%%%%%%%%%%%%%%%%%%%%%%%%%%%%%%%%%%%%%%%%%
\section*{Different studies measure different cognitive systems}
%%%%%%%%%%%%%%%%%%%%%%%%%%%%%%%%%%%%%%%%%%%%%%%%%%%%%%%%%%%%%%%%

To examine how these arguments operate in current Human-AI research, we reviewed a theory-driven sample of studies that differ in their unit of analysis, interaction configuration, and temporal horizon. The aim is not to estimate the prevalence of particular effects or provide an exhaustive review of Human-AI evaluation. Instead, the studies summarized in Table:~\ref{tab:human_ai_evidence}; are used to demonstrate how apparent similarities in claims about the effect of AI may refer to different analytical systems and therefore support contrasting inferences.

At one end of the spectrum, model-centered benchmarks evaluate  $A$ alone ~\cite{ni_survey_2025,srivastava_beyond_2023}. Human performance may provide a comparison baseline \cite{hernandez-orallo_enhancement_2025}, but humans do {\it not} participate in the system being evaluated. Such studies therefore characterize artificial capability rather than Human-AI cognition. Their findings cannot establish whether interaction with the model augments, alters, or impairs human cognition.

A second class of studies evaluates performance while AI assistance is available. These experiments establish whether access to AI changes task performance under coupled conditions~\cite{noy_experimental_2023,dellacqua_navigating_2026}. For example, improvements in writing quality or completion time demonstrate an advantage associated with AI-assisted work. However, unless participants are subsequently assesed without AI, such findigs cannot establish that autonomous human capability is modified.

Other studies explicitly compare the human, AI, and combined configurations~\cite{vaccaro_when_2024, riedl_quantifying_2025}. These designs permit a stronger system-level question: does $H+A$ provide capabilities or performance advantages relative to its components? Such comparisons distinguish simple human augmentation from synergy relative to the best stand-alone model. They can therefore identify properties specifically associated with the combined configuration, although they do not by themselves establish constitutive cognitive integration.

A separate question is addressed when the human is reevaluated after AI assistance is withdrawn. Such designs distinguish improved performance of the coupled or assisted configuration from changes in autonomous human capability. Wu {\it et al.}~\cite{wu_human-generative_2025}, for instance, report substantial benefits during collaboration but {\it limited} transfer to subsequent solo performance, whereas Str{\"o}mberg {\it et al.}~\cite{stromberg_generative_2026} examined improved AI-assisted homework outcomes alongside poorer subsequent closed-book learning outcomes. These results are not incompatible with findings of contemporaneous performance enhancement: they are about different systems at different times.

Earlier works also show that the direction of an effect cannot be inferred from the mere presence of AI. Outcomes vary with task demands, human and artificial capabilities, and the architecture through which their contributions are coordinated~\cite{wang_exploring_2025,singh_systematic_2025,davis_interaction-centered_2026,dellacqua_navigating_2026,li_human-ai_2026}. The same AI system can support function on tasks within its capability frontier while degrading performance when its outputs are unreliable or poorly matched to the job. Likewise, alternative collaboration protocols can produce inconsistent outcomes even when the underlying human and AI components remain unchanged.

The resulting literature is therefore less contradictory than it initially appears~\cite{xu_applying_2024,zhao_understanding_2026,hernandez-orallo_enhancement_2025}. Studies frequently answer different questions: how capable is $A$; how well does a human perform with assistance; what can $H+A$ accomplish as a coupled configuration; or how does $H_{\rm post}$ perform after decoupling? These outcomes are not interchangeable. A cumulative science of Human-AI cognition consequently requires that findings be indexed to both the cognitive system evaluated and the coupling configuration under which the effect was observed.

%%%%%%%%%%%%%%%%%%%%%%%%%%%%%%%%%%%%%%%%%%%%%%%%%%%%%%%%%%%%%%%%%%%
\section*{The Human-LLM Cognitive Extension framework ({\tt HLCE})}
%%%%%%%%%%%%%%%%%%%%%%%%%%%%%%%%%%%%%%%%%%%%%%%%%%%%%%%%%%%%%%%%%%%

The preceding analysis suggests that research on Human-LLM cognition requires a methodological framework~\cite{xu_applying_2024,zhao_understanding_2026,heersmink_dimensions_2015} that separates three questions that are often conflated: which system is being evaluated, how that system is coupled, and what inference the resulting evidence can support. The {\tt HCLE} framework is proposed as an architecture for making these distinctions explicit. Notably, {\tt HCLE} does not assume that a Human-LLM configuration constitutes an extended cognitive system. Instead, it specifies the analytical and experimental steps required to move from observations of assisted performance toward progressively stronger claims about system-level capability, human transformation, and, where relevant, constitutive cognitive integration.

At time \(t\), the coupled Human-LLM configuration is represented as
%%%%%%%%%%%%%%%%%%%%%%%%%%%%%%%%%%%%%%%%%%
\begin{equation}
\System(t;\Coupling) = \textcolor{black}f\bigl( \Hsys(t), \Asys(t), I(t;\Coupling) \bigr),
\label{eq:hlce_system}
\end{equation}
%%%%%%%%%%%%%%%%%%%%%%%%%%%%%%%%%%%%%%%%%%
where $\Hsys(t)$ denotes the task-relevant state of the human component, $\Asys(t)$ the corresponding state of the artificial component, $I(t;\Coupling)$ the evolving interaction, and $\Coupling$ the coupling structure through which interaction is organized. The function $f$ conceptually represents the dependence of the coupled configuration on these components and their interaction, without prescribing a particular functional form or assuming constitutive cognitive integration. This representation is intentionally minimal. Its purpose is not to provide a complete cognitive model or predict the dynamics of the coupled system, but to emphasize that the experimental object depends not only on the human and artificial components, but also on how their interaction is organized. In this sense, ``AI use'' should not be treated as a single homogeneous experimental condition.

{\tt HLCE} can therefore be implemented as a sequence of four methodological decisions: (1) individuate the system to which the outcome belongs; (2) specify the coupling architecture; (3) match the measurement and comparison to the substantive claim; and (4), only when cognitive
integration itself is the hypothesis, apply causal tests appropriate to claims about constitutive participation.

%------------------------------------------------------
\paragraph*{Step 1: Individuate the system under study.}
%------------------------------------------------------
The first step is to specify the system to which the dependent variable is attributed. As we discussed previously, {\tt HLCE} distinguishes at least four analytically different systems $\Hsys,\Asys,\HA,$ and $\Hpost$. Note also that where {\it longitudinal change} is central, $\Hpost$ should be interpreted relative to an explicitly defined baseline such as $\Hpre$, i.e., human before interacting with the artificial system.

The distinction is inferential rather than merely terminological. A finding that the coupled configuration performance (${\cal P}$) is better than the human's performance alone, ${\cal P}(\HA) > {\cal P}(\Hsys)$, supports a claim about assisted or system-level performance. It does not establish that autonomous human capability has improved~\cite{wu_human-generative_2025,stromberg_generative_2026}. That stronger assertion requires evidence such as ${\cal P}(\Hpost) > {\cal P}(\Hpre)$, ideally under a design that controls for practice, repeated testing, selection, and other alternative explanations.

Likewise, a finding that ${\cal P}(\HA) > \max\{{\cal P}(\Hsys),{\cal P}(\Asys)\}$ may support a declaration of system-level advantage or synergy under the chosen operational definition, but it does not by itself imply that the human has learned, that the LLM has become constitutively integrated into human cognition, or that a novel cognitive mechanism has emerged~\cite{vaccaro_when_2024,rosas_reconciling_2020}. {\tt HLCE} therefore requires authors to state explicitly both the system to which each outcome is attributed and the stronger conclusions that the design does {\it not} warrant. 

Several studies report gains in performance during AI-assisted work without equivalent improvements in subsequent unaided performance; the magnitude and direction of post-use effects depend on the task, population, and assessment design~\cite{wu_human-generative_2025,stromberg_generative_2026}.

% ------------------------------------------------------------------
\paragraph*{Step 2: Specify the coupling architecture}
% ------------------------------------------------------------------

Identifying $\HA$ as the relevant system is not sufficient because $\HA$ is not a single experimental condition. The same human, the same LLM, and the same task can produce different outcomes under different interaction compositions. {\tt HLCE} therefore treats coupling $\Coupling$ as part of the experimental object rather than as a fixed background feature.

Building on multidimensional accounts of cognitive integration coupling can be characterized through properties such as information flow, initiative, transparency, reliability, persistence, trust, individualization, and transformation, e.g~\cite{heersmink_dimensions_2015}. These dimensions should not, however, be treated as methodologically equivalent. Within {\tt HLCE} they can play different roles depending on the research question (Table~\ref{tab:hlce_coupling}).

Indeed, recent analyses of AI-mediated cognitive extension have emphasized heterogeneous coupling configurations and their temporal dynamics~\cite{cassinadri_varieties_2026}\cite{chiriatti_system_2025}. HLCE uses these contributions to motivate experimental specification of coupling conditions rather than to infer integration from tool utilization.

%%%%%%%%%%%%%%%%%%%%%%%%%

\newcommand{\rev}[1]{\textcolor{green!50!black}{#1}}
\begin{table}[htbp]
\centering
\small
\renewcommand{\arraystretch}{1.15}

\begin{tabularx}{\linewidth}{
  @{}
  >{\raggedright\arraybackslash}p{0.24\linewidth}
  >{\raggedright\arraybackslash}p{0.28\linewidth}
  >{\raggedright\arraybackslash}X
  @{}
}
\hline
\textbf{Coupling property} &
\textbf{Typical methodological role} &
\textbf{Illustrative operationalization} \\
\hline

Information flow &
Manipulable design variable &
Restrict, expand, sequence, or asymmetrically distribute
the information exchanged between human and LLM. \\

Transparency &
Manipulable design variable &
Compare answer-only output with access to confidence,
evidence, uncertainty, or intermediate rationale
representations. \\

Initiative/control &
Manipulable design variable &
Vary whether the human, LLM, or both may initiate
subgoals, request information, or revise a plan. \\

Reliability &
Manipulable design variable or property of the coupling &
Vary model error rates, calibration, consistency,
or availability across experimental conditions. \\

Durability/persistence &
Manipulable design variable or longitudinal property &
Distinguish persistent system memory from the
duration of human-LLM interaction; compare one-off,
repeated, and longitudinal use. \\

Individualization &
Manipulable design variable or property of the coupling &
Compare generic LLMs with systems adapted to
user-specific history, preferences, or competence.\\

Trust &
State, mediator, or outcome &
Distinguish self-reported trust from behavioral
reliance; assess whether trust mediates delegation
or dependence. \\

Transformation &
Longitudinal outcome &
Assess whether sustained interaction alters human
strategies, skills, representations, or dependence
after decoupling. \\

\hline
\end{tabularx}
\caption{\small Illustrative methodological roles of coupling dimensions within HLCE \cite{heersmink_dimensions_2015}. \textcolor{gray}{Methodological roles are context-dependent; coupling properties may be experimentally manipulated or measured depending on the research design.}}
\label{tab:hlce_coupling}
\end{table}
%%%%%%%%%%%%%%%%%%%%%%%%%

%%%%%%%%%%%%%%%%%%%%%%%%%
The methodological advantage of treating $\Coupling$ explicitly is that the coupling architecture can itself become an experimentally {\it manipulated} variable. A strong design may therefore hold the human, model, task $T$, and evidence constant while changing only selected coupling properties:
%%%%%%%%%%%%
\begin{equation}
\Hsys,\Asys,T = \mathrm{constant},
\qquad
\Coupling_1 \neq \Coupling_2.
\label{eq:coupling_manipulation}
\end{equation}
%%%%%%%%%%%%
Differences in observed performance or interaction dynamics can then be attributed more directly to the organization of the Human-LLM configuration rather than to unspecified differences in ``AI use''. This distinction is especially important because coupling dimensions are not themselves evidence of cognitive extension. High transparency, strong information flow, personalization, or repeated interaction may increase functional integration, but the theoretical significance of those changes depends on the hypothesis being tested and on the evidence used to support it.

% ------------------------------------------------------
\paragraph*{Step 3: Match the measurement to the claim.}
% ------------------------------------------------------

The third step is to specify what evidence would be required by the substantive claim. {\tt HLCE} treats measurement quality as a cross-cutting
requirement, but it does {\it not} assume that a single metric can represent Human-AI cognition across levels of analysis. Reliability, validity, sensitivity, calibration, invariance, and generalization~\cite{pellert_ai_2024}must be evaluated with respect to the construct and system actually being measured. Table~\ref{tab:hlce_claims_evidence}  summarizes the minimum evidential logic for several common claims.

%%%%%%%%%%%%%%%%%%%%%%%%%
\begin{table}[htbp]
\centering
\small
\renewcommand{\arraystretch}{1.2}

\begin{tabularx}{\linewidth}{
@{}
>{\raggedright\arraybackslash}p{0.27\linewidth}
>{\raggedright\arraybackslash}p{0.36\linewidth}
>{\raggedright\arraybackslash}X
@{}
}
\hline
\textbf{Claim} &
\textbf{Minimum comparison or evidence} &
\textbf{What this does not establish by itself} \\
\hline

AI-assisted augmentation &
Assisted human performance versus an appropriate human-only baseline &
Durable human learning, synergy, or cognitive integration. \\

System-level advantage &
$H+A$ versus $H$ &
Advantage over the AI alone or durable change in the human. \\

Synergy/complementarity &
$H+A$ versus both $H$ and $A$, with an explicit synergy definition
and matched task conditions&
Constitutive cognitive integration or human learning. \\

Human learning/transformation &
$H_{\mathrm{post}}$ versus $H_{\mathrm{pre}}$,
ideally with controls and delayed follow-up
to distinguish learning from practice effects &
That the coupled system itself is cognitively extended. \\

Coupling effect &
$C_1$ versus $C_2$, while relevant components and task conditions are controlled
and the manipulated coupling properties are explicitly specified &
That the better-performing architecture is constitutively integrated. \\

Constitutive integration &
Targeted intervention and counterfactual evidence bearing on the causal organization of the mechanism
and distinguishing constitutive participation from external causal support &
Constitutive integration solely from improved performance, reciprocal influence, or intervention effects.\\

\hline
\end{tabularx}

\caption{\small Matching common Human--AI claims to the minimum evidence they require. \textcolor{gray}{The comparisons indicate necessary evidential distinctions, not sufficient conditions for validating stronger cognitive claims.}}
\label{tab:hlce_claims_evidence}

\end{table}

%%%%%%%%%%%%%%%%%%%%%%%%%

This matching step is central because identical numerical outcomes may support different interpretations depending on where the construct is instantiated~\cite{riedl_quantifying_2025,ben-michael_does_2024}. For example, task accuracy measured while assistance is available may characterize an assisted human or a coupled system, depending on the design and theoretical target. It should not be reinterpreted retrospectively as a measure of autonomous human capability. Likewise, post-decoupling tests provide evidence about the human after interaction but not necessarily about the structure of the coupled system that preceded them.

%\textcolor{myorange}{Psychometric methods \cite{pellert_ai_2024} can complement performance-oriented evaluations within {\tt HLCE} by providing established approaches to construct operationalization, measurement reliability, and validity assessment. Pellert et al.  illustrate how psychometric inventories can be adapted to characterize LLM behavior, while highlighting the methodological challenges of transferring human-oriented instruments to artificial systems. Within {\tt HLCE}, psychometric assessment must therefore distinguish between properties of the human participant, the LLM, and the coupled system, ensuring that each construct is explicitly defined and that its measurement is validated for the intended level of analysis.}

%Recent applications to LLMs have demonstrated the use of established psychometric inventories to characterize behavioral properties such as personality-related traits, values, and attitudes~\cite{pellert_ai_2024}. However, measures developed for individual models cannot be assumed to generalize to integrated Human--LLM configurations. From an {\tt HLCE} perspective, the relevant constructs, measurement procedures, and validation criteria must therefore be aligned with the unit of analysis under investigation.

Psychometric methods can complement performance-oriented evaluations within {\tt HLCE} by providing established approaches to construct operationalization, measurement reliability, and validity assessment. Recent applications to LLMs have demonstrated the use of established psychometric inventories to characterize behavioral properties such as personality-related traits, values, and attitudes~\cite{pellert_ai_2024}. Such applications also raise methodological questions about transferring human-oriented psychological constructs and instruments to artificial systems. However, measures developed for individual models cannot be assumed to generalize to integrated Human--LLM configurations. From an {\tt HLCE} perspective, the relevant constructs, measurement procedures, and validation criteria must therefore be aligned with the unit of analysis under investigation.

% ------------------------------------------------------------------
\paragraph*{Step 4: Test integration causally when the hypothesis requires it.}
% ------------------------------------------------------------------
This step applies only when the theoretical claim concerns the boundaries of cognition. If the research question is whether one interface, prompting regime, or collaboration protocol improves performance, no claim about extended cognition is required. If the hypothesis is instead that the LLM participates constitutively in the mechanism realizing the cognitive process, stronger causal evidence is needed.

Following mechanistic approaches to cognitive boundaries~\cite{kaplan_how_2012}, {\tt HLCE} treats intervention as a way to test whether the artificial component is merely causally relevant or {\it participates} more deeply in the mechanism responsible for the target phenomenon. Importantly, the following notions should remain distinct: {\tt causal influence} $\neq$ {\tt reciprocal influence} $\neq$ {\tt mutual manipulability} $\neq$ {\tt constitutive integration}.

A message from the LLM that changes the human's belief state provides evidence of causal influence. Bidirectional effects across turns may
provide evidence of reciprocal dependence. Stronger claims require a design in which interventions on candidate components systematically alter the behavior of the mechanism and in which the causal role of those components can be distinguished from that of external enabling conditions. Even then, mutual manipulability should be treated as evidence compatible with constitutive integration rather than as an automatic confirmation that a Human-LLM configuration constitutes a single cognitive system.

{\tt HLCE} therefore places claims about cognitive extension \cite{heersmink_dimensions_2015,kaplan_how_2012} at the end of an evidential sequence rather than at the beginning of the analysis (Figure~\ref{fig:hlce_workflow}): {\tt assistance} $\rightarrow$ {\tt reciprocal coordination} $\rightarrow$ {\tt functional integration} $\rightarrow$ {\tt candidate constitutive integration}. Each transition requires additional evidence. In this sense, {\tt HLCE} is designed to constrain interpretation while expanding the set of phenomena that can be measured.

%%%%%%%%%%%%%%%%%%%%%%%%%%%%%%%%%%%%%%%%%%%%%%%%%%%%%%%%%%%%%%%%%%%%%%%%%%%%%
\begin{figure}[htbp]
\centering
\small

% STEP 1
\fcolorbox{black!60}{black!9}{%
\parbox{0.88\linewidth}{%
\centering
\textbf{1. INDIVIDUATE THE COGNITIVE SYSTEM}\\[5pt]
$H \quad|\quad A \quad|\quad H{+}A
\quad|\quad H_{\mathrm{post}}$
}}

\par\vspace{7pt}
{\large $\downarrow$}
\par\vspace{7pt}

% STEP 2
\fcolorbox{black!60}{black!9}{%
\parbox{0.88\linewidth}{%
\centering
\textbf{2. SPECIFY THE COUPLING ARCHITECTURE}\\[5pt]
Manipulate $C$ under controlled conditions:
$C_1 \neq C_2$
}}

\par\vspace{7pt}
{\large $\downarrow$}
\par\vspace{7pt}

% STEP 3
\fcolorbox{black!60}{black!9}{%
\parbox{0.88\linewidth}{%
\centering
\textbf{3. ALIGN EVIDENCE WITH CLAIMS}\\[5pt]
Assisted performance $\;|\;$ System-level advantage\\
Synergy $\;|\;$ Post-interaction change
}}

\par\vspace{7pt}
{\large $\downarrow$}
\par\vspace{7pt}

% STEP 4 - OPTIONAL
\fcolorbox{black!60}{black!4}{%
\parbox{0.88\linewidth}{%
\centering
\textbf{4. OPTIONAL: TEST COGNITIVE INTEGRATION}\\[5pt]
Targeted interventions and counterfactual
evidence of mechanistic participation
}}

% SPACE BEFORE INFERENTIAL CONSTRAINT
\par\vspace{14pt}

% INFERENTIAL CONSTRAINT
\par\vspace{14pt}

{\centering
\footnotesize
\textbf{INFERENTIAL CONSTRAINT:}
Improved performance or reciprocal influence does not,
by itself, establish constitutive cognitive integration.
\par}

\par\vspace{18pt}
\caption{ \small HLCE as an experimental and inferential strategy. \textcolor{gray}{The framework specifies the unit of analysis, manipulates coupling conditions, and aligns evidence with the claims it can support. Causal tests of constitutive cognitive integration are optional and require additional mechanistic evidence. HLCE does not assume that Human--LLM configurations constitute extended cognitive systems.}}
\label{fig:hlce_workflow}
\end{figure}

%%%%%%%%%%%%%%%%%%%%%%%%%%%%%%%%%%%%%%%%%%%%%%%%%%%%
\section*{Analytical tools compatible with {\tt HLCE}}
%%%%%%%%%%%%%%%%%%%%%%%%%%%%%%%%%%%%%%%%%%%%%%%%%%%%

{\tt HLCE} does not prescribe a single analytical technique. Once the system, coupling architecture, construct, and intended inference have been specified, different methods can be selected according to the research question. Table~\ref{tab:analytic_tools} summarizes these analytical techniques in relation to the selected research question. Graph-based models, semantic embeddings, state-space or partially observable decision models, information-theoretic measures, and counterfactual interventions are therefore best treated as {\it analytical tools compatible with {\tt HLCE}}, rather than as constitutive dimensions of the framework itself. 

This distinction is important because analytic methods operate at different explanatory levels. A temporal graph can characterize the
organization of interaction; an embedding model can quantify representational change; a Partially Observable Markov Decision Process (POMDP) type model can formalize sequential decision making under partial observability; compression-based statistics can describe structural regularity; and counterfactual perturbations can estimate causal influence. We review these methods below, but note that none of them, taken alone, establishes that a Human-LLM system is cognitively extended.

%%%%%%%%%%%%%%%%%%%%%%%%%%%%%%%%%%%%%%%%%%%%
\begin{table}[htbp]
\centering
\begin{tabularx}{\textwidth}{>{\raggedright\arraybackslash}p{0.23\textwidth}
                            >{\raggedright\arraybackslash}p{0.24\textwidth}
                            >{\raggedright\arraybackslash}p{0.25\textwidth}
                            X}
\toprule
\textbf{Research question} & \textbf{Candidate method} & \textbf{What it can characterize} & \textbf{What it cannot establish alone} \\
\midrule
How are control, evidence, or information distributed over time? &
Temporal/dynamic graphs \cite{holme_temporal_2012} &
Connectivity, centrality, reciprocity, dependency paths, temporal organization &
Cognitive extension or epistemic correctness \\

Do the human and LLM representations converge, diverge, or explore new regions? &
Embedding-based trajectory analysis \cite{reimers_sentence-bert_2019}&
Semantic distance, convergence, divergence, clustering, representational shifts &
Correctness, causal dependence, or constitutive integration \\

How do agents act under partial observability and update beliefs over time? &
State-space or POMDP-type model \cite{kaelbling_planning_1998} &
Latent states, observations, actions, transition dynamics, policy adaptation &
That two agents form a single cognitive mechanism \\

How structurally repetitive or compressible is the interaction? &
Compression/ description-length measures ~\cite{cilibrasi_clustering_2004} &
Redundancy, predictability, structural regularity &
Cognitive novelty, epistemic value, or emergence \\

Does a contribution from one component alter the state or action of the other? &
Intervention/ counterfactual analysis \cite{kaplan_how_2012} &
Direction and magnitude of causal influence &
Mutual manipulability or constitutive integration without stronger tests \\

Does the coupled system produce solutions not reducible to either initial component baseline? &
Baseline comparison, novelty analysis, solution-space analysis ~\cite{rosas_reconciling_2020}&
Candidate complementarity, novelty, or non-reducibility &
Formal cognitive emergence without an explicit emergence criterion \\
\bottomrule
\end{tabularx}
\caption{\small \textcolor{gray}{Illustrative analytical tools that can be selected after the {\tt HLCE} research question has been specified.}}
\label{tab:analytic_tools}
\end{table}
%%%%%%%%%%%%%%%%%%%%%%%%%%%%%%%%%%%%%%%%%%%%%%%

%--------------------------------
\paragraph{Process organization.}
%--------------------------------
Temporal or graph-based representations \cite{holme_temporal_2012} can be useful when the research question concerns how information, evidence, or control propagate through the interaction. Nodes may represent claims, actions, evidence, or task states, while timestamped edges represent dependency or information transfer. Measures such as centrality, reciprocity, modularity, path length, or temporal reachability can then describe how the organization of $\HA$ changes over time. These descriptors are particularly useful when two coupling architectures achieve similar final performance through different interaction processes.

%--------------------------------
\paragraph{Representational dynamics.}
%--------------------------------
Embedding-based analyses \cite{reimers_sentence-bert_2019} can be used when the research question concerns the evolution of semantic content. Human and LLM contributions may be projected into a common representational space and compared through semantic distance, convergence, divergence, clustering, or trajectory similarity. Such analyses can reveal whether the presence of interaction produces increasing alignment, sustained disagreement, or movement toward representations not present in the initial contributions. However, semantic convergence should not be equated with epistemic correctness, and representational novelty should not be equated with cognitive emergence.

%--------------------------------
\paragraph{Sequential decision models.}
%--------------------------------
When interaction is modeled as a sequential decision process, state-space models or partially observable Markov decision processes (POMDPs) \cite{kaelbling_planning_1998} may provide a useful formal language. Such models can represent {\it latent} task states, observations, actions, uncertainty, belief updates, and policies. Their role in {\tt HLCE} is generative and explanatory: they can specify how a human and an artificial agent update their states in response to partial information. A standard POMDP does not, however, automatically capture two mutually adapting agents or establish that the resulting configuration constitutes a single cognitive system. Any such claim requires an explicit mapping between the formal model and the theoretical construct being tested.

%--------------------------------
\paragraph{Structural complexity.}
%--------------------------------
Compression-based measures \cite{cilibrasi_clustering_2004} can provide descriptive information about the structural regularity of interaction trajectories. Measures of compressibility, description length, or redundancy may distinguish a highly repetitive exchange from one with greater structural variety. These measures should be interpreted cautiously because lower compressibility can also arise from noise, verbosity, or superficial lexical variation. Within {\tt HLCE}, they are therefore exploratory descriptors that should be interpreted alongside independent measures of task performance, semantic structure, or causal organization.

%--------------------------------
\paragraph{Intervention and counterfactual analysis.}
%--------------------------------
Intervention-based methods \cite{kaplan_how_2012} are most directly relevant when the research question concerns causal influence across the boundaries of the system. Examples include withholding a message, perturbing available evidence, altering confidence information, changing the order of contributions, or selectively disabling a communication channel. Comparing observed trajectories with counterfactual or perturbed trajectories can estimate whether one component's contribution changes the state or action of the other. The interpretation should remain appropriately limited: message-level influence is evidence of causal dependence, but not, by itself, of mutual manipulability or constitutive integration.

%--------------------------------
\paragraph{Complementarity, novelty and emergence analysis.}
%--------------------------------
Baseline comparisons and solution-space analyses can be used to examine whether Human–LLM interaction produces outcomes that differ from or exceed those achievable by either component independently. Such analyses may characterize performance complementarity, solution novelty, or candidate non-reducibility. Information-theoretic approaches to causal emergence\cite{rosas_reconciling_2020}  provide a more formal perspective by examining whether collective system properties exhibit dynamically relevant information that cannot be attributed to individual components alone. However, superior joint performance or novel solutions do not, by themselves, establish causal emergence. Within HLCE, these measures should therefore be treated as exploratory indicators unless explicit emergence criteria, appropriate component-level baselines, and sufficient evidence of the system's causal dynamics are available.
 
%--------------------------------
\subsection*{Why {\tt HLCE} does not require a unified index.}
%--------------------------------
The present version of {\tt HLCE} does not require the proposed dimensions or analytic outputs to be collapsed into a single composite score. Performance, convergence, reciprocity, semantic change, calibration, post-decoupling transfer, and causal dependence refer to different constructs and may vary independently. A unified index would therefore require an explicit measurement model, a principled weighting scheme, and empirical validation demonstrating that the combined score has a stable and interpretable meaning across tasks and coupling architectures. Until those conditions are met, {\tt HLCE} is better understood as a structured experimental architecture than as a scalar measure of ``cognitive extension.''

%%%%%%%%%%%%%%%%%%%%%%%%%%%%%%%%%%%%%%%%%%%%%%%%%%%%%%%%%%%%%%%%%%%%%%%%%%%%%%%%%%%%
\section*{Worked example: the same diagnosis, different coupling architectures}
%%%%%%%%%%%%%%%%%%%%%%%%%%%%%%%%%%%%%%%%%%%%%%%%%%%%%%%%%%%%%%%%%%%%%%%%%%%%%%%%%%%%

To illustrate how the {\tt HLCE} framework can guide experimental design and interpretation, we consider a {\it synthetic} industrial troubleshooting scenario. The example brings together three elements introduced in the preceding sections: the identification of the system under investigation, the specification of its coupling architecture, and the selection of measurements appropriate to that system\cite{heersmink_dimensions_2015,poelitz_benchmark_2026}.
We therefore conceived an scenario involving two simulated agents working together to solve the same diagnostic problem. One represents a human troubleshooter ($H$), and the other represents an LLM assistant ($A$). The agents communicate through a sequence of messages in which they exchange diagnoses, evidence, and proposed actions. We compare two versions of this interaction that differ in how diagnostic information is communicated. The object of analysis is therefore the interacting human-LLM configuration ($H+A$), rather than either agent considered independently.

Note that this is a constructed methodological demonstration, {\it not} an experiment involving human participants. Both agent roles and their interaction trajectories were simulated using an LLM, as described in Materials and Methods. The example illustrates how different interaction conditions can be analyzed within {\tt HLCE}; it does not provide empirical estimates of their effects.

%%%%%%%%%%%%%%%%%%%%%%%%%%%%%%%%%%%%%%%%%%%%%%%%%%%%%%%%%
\paragraph{The diagnostic problem.}
%%%%%%%%%%%%%%%%%%%%%%%%%%%%%%%%%%%%%%%%%%%%%%%%%%%%%%%%%

Consider a factory producing mechatronic devices that must pass a final quality-control test, known as an End-of-Line (EOL) test, before being released. Shortly after the installation of diagnostic software {\tt version 4.2}, the proportion of devices failing this test increases from 1.1\% to 8.4\%. The troubleshooting task is to identify the most plausible cause of the increase and recommend a safe corrective action.

Three possible explanations are considered: ($h_1$) calibration drift following maintenance; ($h_2$) a mechanical problem involving connector seating or insertion force; and ($h_3$) a software-related problem in which the new version allows insufficient time for the device to respond during testing. The scenario also provides evidence that can help distinguish these explanations. Seven of 30 devices tested with {\tt version 4.2} fail, whereas all 30 pass when the previous software version is restored. The rejected devices also pass an independent bench test, another testing station using {\tt version 4.1} remains stable with the same production batch, and the configured response timeout has been reduced from 180 to 120\,ms.

Taken together, these observations favor a software-related explanation over a persistent hardware defect. For the purposes of the simulation, $h_3$ is therefore defined as the correct diagnosis. The target corrective action ($A_3$) is to contain potentially affected units, temporarily restore {\tt version 4.1}, conduct a controlled comparison, and correct the configuration of {\tt version 4.2} before redeployment.

%%%%%%%%%%%%%%%%%%%%%%%%%%%%%%%%%%%%%%%%%%%%%%%%%%%%%%%%%
\paragraph{The two simulated agents and their interaction.}
%%%%%%%%%%%%%%%%%%%%%%%%%%%%%%%%%%%%%%%%%%%%%%%%%%%%%%%%%

The example begins with a disagreement between the two agents. The {\it simulated} human troubleshooter initially favors calibration drift ($h_1$), with an assigned confidence of 0.55. The {\it simulated} LLM favors the software timeout explanation ($h_3$), with an assigned confidence of 0.87. These values define the agents' initial states within the simulation; they are not measured human judgments or independently validated confidence estimates.

The agents then engage in a dialogue to resolve the problem. During this exchange, the ``human'' can question the LLM's recommendations, request supporting information, and reconsider the diagnosis. The LLM can propose explanations, communicate relevant evidence, and help develop a corrective action. Each message can therefore influence the information available to the other agent and the subsequent direction of the investigation.

This interaction is the central feature of the example. The diagnostic outcome is not considered solely as the response of the human or the LLM, but as the result of an exchange between them. In the terminology of {\tt HLCE}, the system being examined is $H+A$, and the organization of their communication is part of its coupling form \Coupling.

%%%%%%%%%%%%%%%%%%%%%%%%%%%%%%%%%%%%%%%%%%%%%%%%%%%%%%%%%
\paragraph{The same problem, two communication conditions.}
%%%%%%%%%%%%%%%%%%%%%%%%%%%%%%%%%%%%%%%%%%%%%%%%%%%%%%%%%

We construct two ten-turn dialogues involving the same simulated roles, initial diagnoses, diagnostic problem, and available evidence. The difference lies in what information the LLM communicates at the beginning of the exchange.

In the {\tt low}-informational-transparency condition (LOW\_IT), the LLM initially communicates its preferred diagnosis ($h_3$) and confidence, but provides little supporting evidence. The simulated human must therefore request clarification and supporting observations before reconsidering its initial diagnosis. Moreover, in the {\tt high}-informational-transparency condition (HIGH\_IT), the LLM communicates its preferred diagnosis together with the evidence supporting it, alternative explanations, and remaining uncertainty. The human can consequently examine the reasoning behind the recommendation earlier in the dialogue.

The two conditions thus represent alternative ways of organizing information exchange between the same two agent roles. They illustrate a manipulation of coupling architecture rather than a change in the diagnostic task or the predefined correct answer. Importantly, they differ in the timing and amount of evidence communicated, not only in the transparency of its presentation.

%%%%%%%%%%%%%%%%%%%%%%%%%%%%%%%%%%%%%%%%%%%%%%%%%%%%%%%%%
\paragraph{Different interactions, the same final diagnosis.}
%%%%%%%%%%%%%%%%%%%%%%%%%%%%%%%%%%%%%%%%%%%%%%%%%%%%%%%%%

\begin{figure}[H]
\centering
\includegraphics[width=\linewidth]{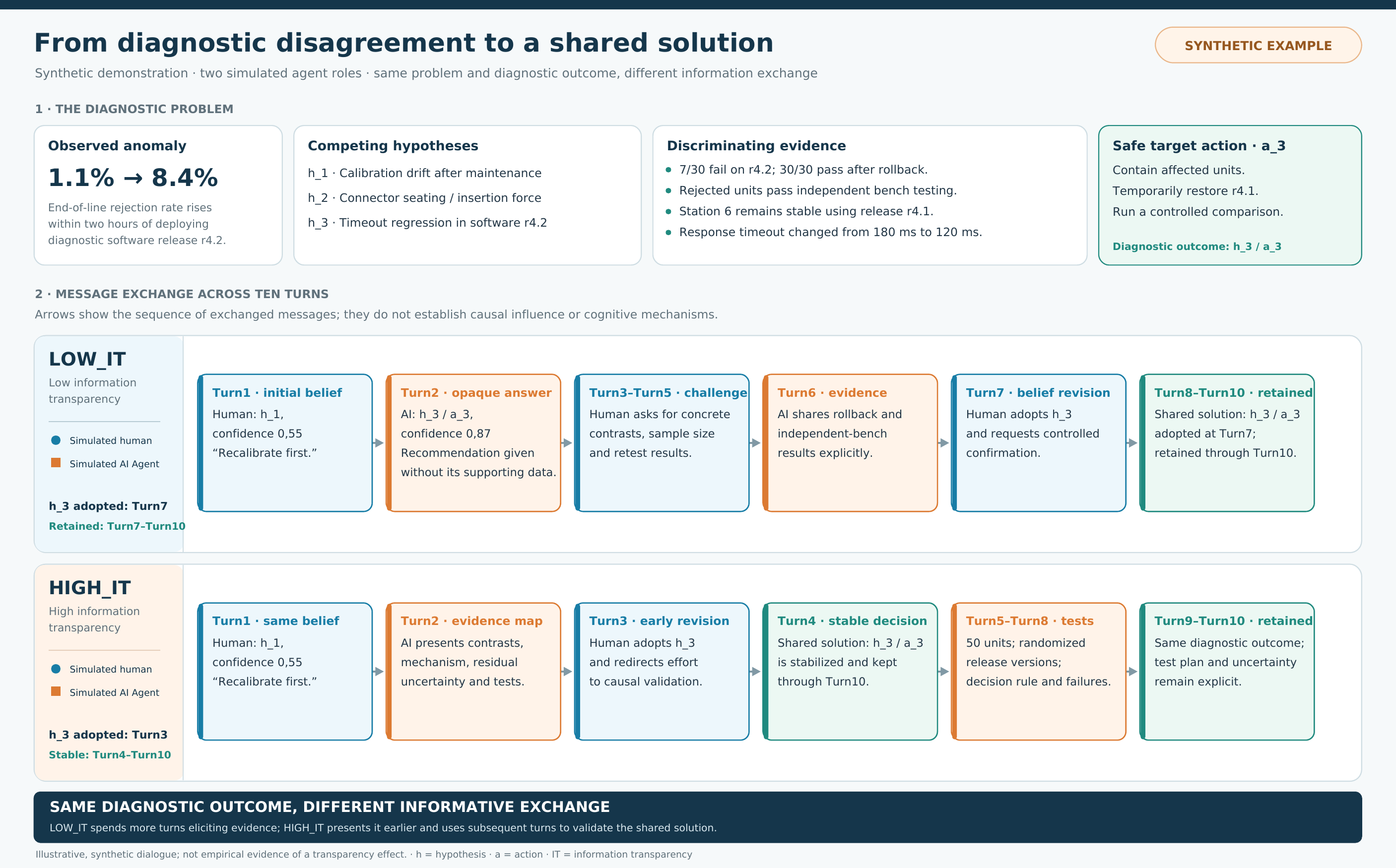}
\caption{\small \textbf{Same diagnostic problem, different interaction trajectories.}\textcolor{gray}{ Two simulated agents, representing a ``human'' troubleshooter ($H$) and an LLM assistant ($A$), exchange messages to diagnose an increase in End-of-Line (EOL) test failures following the introduction of a new software version. Both dialogues start from the same disagreement and ultimately identify the predefined correct diagnosis ($h_3$) and corrective action ($A_3$). In LOW\_IT, supporting evidence is communicated later, following requests for justification. In HIGH\_IT, the LLM makes this evidence available {\it earlier}, allowing the dialogue to move sooner toward corrective-action planning. The figure represents two scripted trajectories, not observed human--LLM interactions or estimated effects of informational transparency.}}
\label{fig:hlce_interaction}
\end{figure}

Figure~\ref{fig:hlce_interaction} summarizes the two simulated dialogues. Under LOW\_IT, the LLM recommends $h_3$ at turn 2, but the human does not immediately accept it. The following exchanges focus on questioning the recommendation and obtaining supporting evidence. The rollback and independent-bench observations become explicit at turn 6, after which the human adopts $h_3$ at turn 7. The dialogue subsequently develops the shared corrective action.

Under HIGH\_IT, the LLM communicates the relevant evidence and uncertainty at turn 2. The human adopts $h_3$ at turn 3, allowing the subsequent interaction to focus on the proposed corrective action, including a controlled comparison, rollback procedures, and criteria for evaluating the correction.

Both dialogues continue for ten turns and ultimately reach the same predefined diagnosis ($h_3$) and target action ($A_3$). However, the interactions differ in how the agents arrive at that result: LOW\_IT devotes more of the exchange to obtaining and checking the justification for the recommendation, whereas HIGH\_IT makes this information available earlier and shifts the discussion toward planning and verification.

%%%%%%%%%%%%%%%%%%%%%%%%%%%%%%%%%%%%%%%%%%%%%%%%%%%%%%%%%%%%%%%%%%%%%%%%%%%%%%%%%%%%%%%

%%%%%%%%%%%%%%%%%%%%%%%%%%%%%%%%%%%%%%%%%%%%%%%%%%%
\paragraph{What should be measured?}
%%%%%%%%%%%%%%%%%%%%%%%%%%%%%%%%%%%%%%%%%%%%%%%%%%%
If the two conditions were evaluated solely by final diagnostic accuracy, they would appear equivalent: both reach $h_3$ and $A_3$. Yet the dialogues differ in when the human adopts the correct diagnosis and how the agents use their exchanges. This distinction illustrates why evaluating an interacting system requires measures of the interaction process as well as its final outcome \cite{riedl_quantifying_2025,poelitz_benchmark_2026}.

Table~\ref{tab:hlce_core_metrics} provides a minimal descriptive comparison. The simulated human adopts $h_3$ at turn 7 under LOW\_IT and at turn 3 under HIGH\_IT. The HIGH\_IT dialogue also contains more text, indicating that earlier diagnostic convergence need not correspond to lower communication volume.

%%%%%%%%%%%%%%%%%%%%%%%%%%%%%%%%%%%%%%%%%%%%%%%%%%%
\begin{table}[H]
\centering
\small
\begin{tabularx}{\linewidth}{@{}>{\raggedright\arraybackslash}Xcc@{}}
\toprule
\textbf{Descriptive outcome} & \textbf{LOW\_IT} & \textbf{HIGH\_IT} \\
\midrule
Final diagnosis/action & $h_3/A_3$ & $h_3/A_3$ \\
First human adoption of $h_3$ & Turn 7 & Turn 3 \\
Total dialogue turns & 10 & 10 \\
Dialogue length & 247 tokens & 315 tokens \\
\bottomrule
\end{tabularx}
\caption{\small \textbf{Descriptive comparison of the two synthetic dialogues.} \textcolor{gray}{Turn numbers identify positions within individual scripted dialogues, not averages across repeated simulations. Both conditions reach the same predefined diagnostic endpoint, while differing in the timing of diagnosis adoption and the amount of text exchanged. These quantities describe the constructed trajectories and are not empirical effect estimates.}}
\label{tab:hlce_core_metrics}
\end{table}
%%%%%%%%%%%%%%%%%%%%%%%%%%%%%%%%%%%%%%%%%%%%%%%%%%%

Additional process measures could characterize when evidence is introduced, how diagnoses change during the exchange, and whether the agents reach and maintain agreement. Such measures provide information about the organization of the joint problem-solving process, but do not by themselves establish improved human reasoning, learning, or cognitive integration.

%%%%%%%%%%%%%%%%%%%%%%%%%%%%%%%%%%%%%%%%%%%%%%%%%%%%%%%%%
\paragraph{What does this example illustrate within {\tt HLCE}?}
%%%%%%%%%%%%%%%%%%%%%%%%%%%%%%%%%%%%%%%%%%%%%%%%%%%%%%%%%

The example illustrates the methodological distinctions developed in the preceding sections. First, the unit of analysis is explicitly defined as the interacting configuration, \HA. Second, the coupling is made experimentally relevant by comparing two protocols for communicating diagnostic information. Third, the selected measures distinguish the final outcome from the interaction trajectory leading to it. Finally, the interpretation is restricted to what the simulated comparison can support.

The two dialogues demonstrate that an identical final answer can arise through different patterns of information exchange, questioning, and revision. They do not demonstrate that HIGH\_IT improves human cognition, that the LLM produces a synergistic advantage, or that the two agents constitute an integrated cognitive mechanism. Furthermore, because HIGH\_IT changes both the timing and amount of information available, the example does not isolate the independent causal effect of transparency. The dialogues are constructed rather than observed, and no inferential comparison is possible.

The broader implication is that conclusions about Human--LLM cognition depend on identifying the system being evaluated and measuring the properties relevant to it. In this example, {\tt HLCE} directs attention from the final diagnosis alone to the {\it interaction} through which it is reached. Establishing assisted benefits, durable human learning, synergy, or constitutive integration would require additional experimental conditions, appropriate standalone baselines, and targeted causal tests.

%%%%%%%%%%%%%%%%%%%%%%%%%%%%%%%%%%%%%%%%
\section*{Discussion}
%%%%%%%%%%%%%%%%%%%%%%%%%%%%%%%%%%%%%%%%

Previous studies on whether LLMs benefit or harm cognition have not reached a single conclusion~\cite{noy_experimental_2023,wu_human-generative_2025,stromberg_generative_2026,dellacqua_navigating_2026}. Although this may partly reflect the complexity of their effects, we argue here that some of the apparent disagreement stems from differences in the systems being measured~\cite{hernandez-orallo_enhancement_2025, vaccaro_when_2024}. For instance, assisted performance may improve while the human later shows little transfer, because the execution of $H + A$ and that of $H_{\mathrm{post}}$ correspond to disparate systems~\cite{wu_human-generative_2025,stromberg_generative_2026} Conversely, a decline in unaided operation does not erase a gain obtained by the coupled configuration when the LLM is available~\cite{stromberg_generative_2026}.

{\tt HLCE} offers a way to report such conflicting observations without attributing them to complexity alone or making unspecified claims about ``human cognition''. The four-system indexing convention should therefore be treated as a reporting {\it requirement}, not as an empirical declaration that each of the four systems has the same psychological status \cite{xu_applying_2024,hernandez-orallo_enhancement_2025}.

How a human and an LLM solve a problem can depend on what information they exchange, when they share it, and who takes the initiative to check a proposed answer~\cite{poelitz_benchmark_2026,singh_systematic_2025,li_human-ai_2026}. Our synthetic example illustrates this by comparing an LLM that provides a diagnosis with limited explanation with one that also shares supporting evidence and uncertainty. However, because the second condition makes more information available earlier, the difference cannot be attributed to transparency alone. Future experiments should examine these factors separately, including the amount of evidence shared, the length of messages, the communication of confidence, and who initiates verification \cite{heersmink_dimensions_2015,poelitz_benchmark_2026}. They should also assess not only whether the final answer is correct, but how it was reached: faster agreement may reflect better use of evidence or simply premature acceptance of the LLM's suggestion~\cite{vaccaro_when_2024,dellacqua_navigating_2026}.

Claims that augmentation require an appropriate human-only comparison~\cite{noy_experimental_2023}; statements of synergy require comparison with the best standalone component under a declared performance scale~\cite{vaccaro_when_2024,riedl_quantifying_2025}; allegations that people learn from, become dependent on, or lose skills through LLM use require pre/post assessment, ideally with delayed follow-up~\cite{wu_human-generative_2025,stromberg_generative_2026}. An apparent system-level advantage cannot on its own distinguish complementarity from division of labor, selective delegation, or an increase in information available to the team ~\cite{vaccaro_when_2024,luo_agentds_2026}. Conversely, changes in $H_{\mathrm{post}}$ may occur through learning or changes in motivation and strategy without implying that the prior configuration was constitutively extended ~\cite{wu_human-generative_2025,wilson_re-tooled_2010}. This motivates reporting a set of outcome measures, rather than compressing them into a premature single {\tt HLCE} score.

Beyond these measurement issues, theoretical claims about extended cognition require a different evidential approach from that used in interface comparisons ~\cite{clark_extended_1998,kaplan_how_2012}, A user revising a diagnosis after reading an LLM message may indicate the influence of the information received, but does not by itself establish causality. Bidirectional revisions may suggest a stronger pattern of interdependence, but are still compatible with two independent systems exchanging messages. An argument for constitutive integration would require targeted interventions on putative components and the joint mechanism, carefully specified criteria for integration, and tests that rule out mere common input or external facilitation~\cite{kaplan_how_2012,heersmink_dimensions_2015}. In {\tt HLCE}, mutual manipulability is thus best treated as a mechanism-oriented research strategy, not as a threshold inferred from large performance effects or high agreement scores.

Yet, several limitations follow from these distinctions. Although {\tt HLCE} is grounded in the EM framework, this is not the only theoretical account of Human-AI cognition; perspectives from distributed cognition, joint cognitive systems, hybrid intelligence, and human-computer interaction may provide complementary or competing interpretations~\cite{zhao_understanding_2026,davis_interaction-centered_2026}. Moreover, even evidence compatible with mutual manipulability would not, by itself, constitute proof of cognitive extension~\cite{kaplan_how_2012}. Convergence must also be distinguished from epistemic correctness, since the human and artificial components may converge on the same explanation even when that explanation is incorrect. Similarly, improved joint performance does not establish cognitive emergence: synergy, complementarity, novelty, and emergence remain analytically distinct constructs~\cite{vaccaro_when_2024,rosas_reconciling_2020}. 

Moreover, the synthetic example demonstrates how the {\tt HLCE} architecture can be applied, but it does {\it not} provide empirical validation of the framework or evidence of cognitive integration in human participants. The proposed coupling manipulations and outcome measures therefore require testing in controlled experiments with human participants and different interaction configurations. Likewise, graph-based, semantic, psychometric, and complexity-based analyses should be regarded as candidate tools rather than validated measures of cognitive extension~\cite{holme_temporal_2012,reimers_sentence-bert_2019,pellert_ai_2024,cilibrasi_clustering_2004}; their usefulness and interpretation will depend on the task, coupling architecture, and construct under investigation \cite{pellert_ai_2024,holme_temporal_2012,reimers_sentence-bert_2019}.

In conclusion, asking whether LLMs improve or degrade cognition is insufficiently specified. Human competence, artificial capability, coupled-system performance, and post-interaction human capacity are distinct empirical objects, and their observed performance and relationships may depend on how human and artificial components are combined. Consistent with extended, distributed, ecological, and network-based perspectives on cognition \cite{clark_extended_1998,heersmink_dimensions_2015,hutchins_cognition_1995,bruner_self_2026}, these distinctions emphasize the importance of identifying the relevant cognitive system in relation to its interactions with external resources, without presupposing constitutive integration. 
%{\tt HLCE} provides a methodological structure for specifying these objects, representing coupling architecture as part of the experimental design, and matching evidence to claims about augmentation, synergy, transformation, and cognitive integration. It treats cognitive integration as a hypothesis to be tested rather than an assumption, while allowing coupling dimensions to serve both as descriptors of interaction and as variables for experimental manipulation. 
Therefore, the central question for Human-AI cognition is not simply whether AI changes cognitive behavior, but which system is being evaluated, under what architecture of interaction, and what kind of cognitive claim the observed evidence is sufficient to support.
Finally, while {\tt HLCE} provides a framework for assessing these changes, it does not encompass the broader implications of AI use for psychological well-being, social organization, or cultural practices. These consequences warrant complementary investigation beyond the scope of the present framework.

\section*{Materials and Methods}
%%%%%%%%%%%%%%%%%%%%%%%%%%%%%%%%

%%%%%%%%%%%%%%%%%%%%%%%%%%%%%%%%
\paragraph{Study set.}
%%%%%%%%%%%%%%%%%%%%%%%%%%%%%%%%
The studies considered here constitute an illustrative, theory-driven sample intended to capture contrasting evaluation approaches rather than to provide an exhaustive review of the field. They were used to compare studies assessing $A$, $H$, $\Coupling$, and $\Hpost$. The selection was intended to illustrate differences in units of analysis, experimental conditions, and assessment times, not to estimate the prevalence or size of effects across the literature. No systematic review or meta-analysis was undertaken. Broader accounts of the evaluation landscape and the limitations of task-oriented benchmarking are provided elsewhere~\cite{ni_survey_2025,hernandez-orallo_enhancement_2025}.

%%%%%%%%%%%%%%%%%%%%%%%%%%%%%%%%%%%%%%%%%%%%%%%%%%%%%%%%%%
\paragraph{Synthetic demonstration of the HLCE framework.}
%%%%%%%%%%%%%%%%%%%%%%%%%%%%%%%%%%%%%%%%%%%%%%%%%%%%%%%%%%

OpenAI's {\tt ChatGPT 5.6 Sol} was used to construct two scripted dialogues representing a diagnostic interaction between a simulated human troubleshooter ($H$) and a simulated LLM assistant ($A$). The purpose was to illustrate how the {\tt HLCE} framework can be applied to $\HA$ under alternative $\Coupling$s. No human participants were involved ({\it both roles were simulated}) and the dialogues do not represent observations from independently tested human-LLM teams.

The simulated task concerned an increase in the End-of-Line (EOL) rejection rate of a mechatronic device following the deployment of a {\it new} diagnostic software. Three competing hypotheses were specified: calibration drift ($h_1$), a connector-related mechanical defect ($h_2$), and a software response-timeout regression ($h_3$). The scenario included observations from software rollback, independent bench testing, a second testing station, and a change in the configured response timeout. The software-related hypothesis ($h_3$) and a corresponding containment, rollback, and verification procedure ($A_3$) were predefined as the correct diagnosis and target action.

The simulated agents were assigned initial diagnostic preferences and confidence values. The human role initially favored $h_1$ with confidence 0.55, while the LLM role favored $h_3$ with confidence 0.87. These values were stipulated starting conditions rather than empirically measured or calibrated beliefs. The dialogue was organized as a sequence of exchanges in which the agents could propose diagnoses, request or communicate evidence, reconsider their positions, and formulate corrective actions. The represented belief revisions were part of the constructed dialogue, rather than outcomes generated by an independently validated cognitive model.

We constructed two interaction conditions using the same diagnostic task, simulated agent roles, initial preferences, evidence inventory, and ten-turn dialogue horizon. In LOW\_IT, the LLM initially communicated its preferred diagnosis and confidence without providing the main supporting evidence; additional evidence was introduced in subsequent exchanges. In HIGH\_IT, the LLM communicated its diagnosis together with supporting observations, alternative explanations, and uncertainty earlier in the interaction. Thus, the conditions differed in the timing, amount, and presentation of communicated information. This contrast was intended to illustrate a difference in coupling architecture, not to isolate the causal effect of informational transparency.

In the end, we compared the final diagnosis and corrective action, the first turn at which the simulated human adopted the predefined correct diagnosis, and the total number of dialogue turns and tokens. Turn numbers were assigned within each dialogue, and the reported quantities describe individual scripted trajectories rather than averages across repeated runs. The interaction diagrams and supplementary descriptors were used to illustrate differences in information exchange and diagnostic revision. No participant-level statistical analyses, replicated simulation estimates, or validated measures of cognitive integration were obtained.

%%%%%%%%%%%%%%%%%%%%%%%%%%%%
\section*{Acknowledgements}
%%%%%%%%%%%%%%%%%%%%%%%%%%%%
This work was supported by grant PID2023-151289NB-I00 (JFP) and PID2024-156477NB-C32 (EB) funded by MICIU/AEI/10.13039/501100011033 and by “ERDF/EU”.

%%%%%%%%%%%%%%%%%%%%%%%%%%%%
\section*{Code availability}
%%%%%%%%%%%%%%%%%%%%%%%%%%%%

Code available https://github.com/ErnestoBiempica/Interface-HLCE

%%%%%%%%%%%%%%%%%%%%%%%%%%%%
\section*{AI use statement}
%%%%%%%%%%%%%%%%%%%%%%%%%%%%
OpenAI's {\tt ChatGPT 5.6 Sol} was used during manuscript preparation to support language editing and stylistic revision, and to assist in the design and production of figures. These editorial uses did not generate empirical evidence or determine the scientific conclusions of the study. The authors take full responsibility for the final content of the manuscript.

\noindent
\bibliography{myrefs_JFP_checked}

\clearpage

\section*{Supplementary Material}
\setcounter{table}{0}
\renewcommand{\thetable}{S\arabic{table}}

\setcounter{table}{0}
\renewcommand{\thetable}{S\arabic{table}}

\begin{table}[htbp]
\centering
\small

\renewcommand{\arraystretch}{1.35}
\setlength{\tabcolsep}{8pt}

\begin{tabularx}{\linewidth}{
@{}
>{\raggedright\arraybackslash}p{0.25\linewidth}
>{\raggedright\arraybackslash}X
@{}
}

\toprule

\textbf{Theoretical contribution} &
\textbf{Implications for cognitive extension} \\

\midrule

\textbf{Clark \& Chalmers (1998)} &
\textbf{Parity principle and cognitive extension.}
Cognition need not be bounded by the biological
organism. External resources may become components
of cognitive processes when functionally integrated. \\

\addlinespace[7pt]

\textbf{Wilson (2002)} &
\textbf{Offloading versus constitutive extension.}
External resources may support cognitive performance
without necessarily becoming components of the
cognitive mechanism. \\

\addlinespace[7pt]

\textbf{Wilson (2010)} &
\textbf{Cognitive transformation through tools.}
Sustained engagement with cognitive artifacts may
reorganize cognitive strategies and processes over time. \\

\addlinespace[7pt]

\textbf{Kaplan (2012)} &
\textbf{Mechanistic boundaries and mutual manipulability.}
Interventions and reciprocal causal dependencies
can inform the investigation of cognitive mechanisms,
although causal influence alone does not establish
constitutive integration. \\

\addlinespace[7pt]

\textbf{Heersmink (2015)} &
\textbf{Multidimensional cognitive integration.}
Human--artifact coupling may differ in degree and
configuration across several dimensions, rather
than constituting a simple binary distinction. \\

\midrule

\addlinespace[5pt]

\textbf{Conceptual progression} &
\textit{Where is cognition?}
$\rightarrow$
\textit{When is a tool only support?}
$\rightarrow$
\textit{How do tools reshape cognition?}
$\rightarrow$
\textit{How can boundaries be tested?}
$\rightarrow$
\textit{How strong is the coupling?} \\

\addlinespace[7pt]

\textbf{Integration dimensions} &
Information flow; reliability; durability; trust;
procedural and informational transparency;
individualization; transformation. \\

\bottomrule

\end{tabularx}

% Separation between table and caption
\par\vspace{12pt}

\caption{
\textbf{Conceptual development of the Extended Mind framework.}
Five theoretical contributions motivate distinct questions
about functional roles, causal support, cognitive
transformation, mechanistic intervention, and degrees
of human--artifact coupling. The schematic does not
imply that interaction with an LLM is by itself
constitutive cognitive extension.
}

\label{table:S1}

\end{table}

% Required packages:
% \usepackage{booktabs}
% \usepackage{longtable}
% \usepackage{array}
% \usepackage{ragged2e}
% \usepackage{pdflscape}
% \usepackage{caption}
%
% Optional:
% \usepackage{microtype}

\newcolumntype{P}[1]{>{\RaggedRight\arraybackslash}p{#1}}

\begin{landscape}
\scriptsize
\setlength{\tabcolsep}{3pt}
\renewcommand{\arraystretch}{1.12}

\begin{longtable}{
P{2.7cm}
P{3.2cm}
P{3.3cm}
P{2.2cm}
P{2.4cm}
P{3.2cm}
P{3.5cm}
}
\caption{Comparative analysis of Human--AI studies according to method, unit of analysis, performance effects, human outcomes, and relevance to the Extended Mind framework.}
\label{tab:human_ai_evidence}\\

\toprule
\textbf{Paper} &
\textbf{Method} &
\textbf{Main result} &
\textbf{Unit of analysis} &
\textbf{Performance effect} &
\textbf{Does AI improve or worsen the human?} &
\textbf{Possible Extended Mind relevance} \\
\midrule
\endfirsthead

\multicolumn{7}{c}{\tablename\ \thetable\ -- continued from previous page} \\
\toprule
\textbf{Paper} &
\textbf{Method} &
\textbf{Main result} &
\textbf{Unit of analysis} &
\textbf{Performance effect} &
\textbf{Does AI improve or worsen the human?} &
\textbf{Possible Extended Mind relevance} \\
\midrule
\endhead

\midrule
\multicolumn{7}{r}{Continued on next page} \\
\endfoot

\bottomrule
\endlastfoot

\textbf{Poelitz, Doshi-Velez \& Lindley (2026)}\newline
\emph{A Benchmark to Assess Common Ground in Human-AI Collaboration \cite{poelitz_benchmark_2026}}
&
Confirmatory user study using a repeated collaborative puzzle task with GPT-4.1. Shared-view and non-shared-view conditions. Task success is exact puzzle match; communication effort, referential conventions, and grounding are analysed longitudinally using GEE and mixed-effects models.
&
Shared visual context improves joint puzzle success. Humans reduce communication effort and adapt their referring expressions, including adopting AI-generated identifiers. Puzzle accuracy itself does not show a learning effect across trials.
&
Human--AI interaction / dyad.
&
Positive for coordination and efficiency; no improvement in joint accuracy over repeated trials.
&
\textbf{Not established.} Humans adapt communicatively, but there is no subsequent human-only test. Neither improvement nor deterioration of human capability can be concluded.
&
Relevant to integration, but not evidence of extension. The paper operationalizes iterative coordination, common ground, and mutual adaptation. It does not test the Extended Mind thesis.
\\

\addlinespace

\textbf{Singh, Hindriks, Heylen \& Baraka (2025)}\newline
\emph{A Systematic Review of Human-AI Co-Creativity \cite{singh_systematic_2025}}
&
Systematic review of 62 papers on co-creative systems. Categorizes creative phase/task, AI proactivity, user control, embodiment, and model type. Human--AI co-creativity is characterized by a human and an AI pursuing a shared goal through bidirectional interaction and mutual adaptation.
&
High user control is associated with greater satisfaction, trust, and ownership; adaptive and context-sensitive proactivity can deepen collaboration. The literature remains heterogeneous and often relies on subjective evaluation.
&
Human--AI interaction, aggregated across reviewed systems.
&
Generally identifies conditions associated with better co-creative experience or performance, but no single causal pooled effect.
&
\textbf{Not established.} The review does not demonstrate persistent improvement or deterioration of human creative ability after AI removal.
&
Conceptually relevant. Bidirectional interaction, mutual adaptation, and shared goals resemble stronger notions of cognitive integration, but the paper does not claim that co-creative systems constitute extended minds.
\\

\addlinespace

\textbf{Luo et al. (2026)}\newline
\emph{AgentDS: Benchmarking the Future of Human-AI Collaboration in Domain-Specific Data Science \cite{luo_agentds_2026}}
&
Benchmark of 17 challenges across six domains, with realistic synthetic multimodal datasets, plus a 10-day competition with 29 teams and 80 participants. Competition solutions are compared with AI-only baselines. Human--AI workflows are reconstructed from reports, code, and qualitative analysis.
&
Current autonomous agents struggle with domain-specific reasoning. Successful teams use humans for domain knowledge, hypothesis formation, error diagnosis, and strategic decisions, while AI accelerates coding, experimentation, and iteration. The best Human--AI teams outperform the autonomous baselines.
&
Primarily Human--AI interaction, with AI-only systems as comparison.
&
Positive joint-system performance.
&
\textbf{Not established.} Participants were not randomly assigned to collaboration strategies, and human competence before and after collaboration was not measured. Collaboration analysis is observational.
&
Strong empirical relevance to functional complementarity. Human-initiated feedback loops distribute different cognitive functions across human and AI. This is compatible with analysing the coupled system as a unit, but does not establish constitutive cognitive extension.
\\

\addlinespace

\textbf{Smart, Clowes \& Clark (2025)}\newline
\emph{ChatGPT, Extended: Large Language Models and the Extended Mind \cite{smart_chatgpt_2025}}
&
Technical/philosophical study. \emph{Digital Andy} combines GPT-3.5 with RAG over 58 Clark publications, treating the vector store as non-parametric external memory. RAG and non-RAG configurations are compared.
&
External information influences LLM responses. However, the authors ultimately reject \emph{Digital Andy} as a genuine EXAI system, primarily because the LLM itself does not initiate the retrieval loop; the external application does.
&
Model + external memory, not Human--AI interaction.
&
RAG changes and improves Clark-specific model responses.
&
\textbf{Not measured.} There is no experiment on a human user.
&
Direct Extended Mind paper. It explicitly transfers active externalism to LLM architecture, but does not demonstrate Human--LLM extended cognition. Its negative conclusion makes agency and control part of the boundary problem.
\\

\addlinespace

\textbf{Di Santi (2026)}\newline
\emph{Cognitive Amplification vs Cognitive Delegation in Human-AI Systems \cite{di_santi_cognitive_2026}}
&
Conceptual/mathematical framework plus NetLogo agent-based simulations. Introduces CAI*, Dependency Ratio, Human Reliance Index, and Human Cognitive Drift Rate.
&
Across tested simulations, no configuration achieves genuine amplification: mixed reliance preserves more modeled human capability than full delegation, but $CAI^*<0$ and $D>1$. Even zero atrophy does not produce positive collaborative gain under the modeled dynamics.
&
Simulated Human--AI system.
&
In simulation, acceptable hybrid output can coexist with dependency and negative human drift.
&
\textbf{No empirical conclusion about humans.} Deterioration occurs in the model, not in measured human participants.
&
Directly relevant conceptually. The paper separates joint-system performance from retained autonomous human capability and proposes an operational distinction between amplification and delegation, but remains simulation-based.
\\

\addlinespace

\textbf{Ben-Michael et al. (2025)}\newline
\emph{Does AI Help Humans Make Better Decisions? \cite{ben-michael_does_2024}}
&
Statistical causal-evaluation framework applied to an RCT of pretrial risk-assessment recommendations. Compares human decisions with and without AI recommendations; AI-alone performance requires additional counterfactual and partial-identification treatment because of selective labels.
&
Providing the risk assessment does not improve classification accuracy of judges' cash-bail decisions. AI-alone alternatives considered in the paper do not outperform human decisions.
&
Human decision-making under Human--AI assistance, plus AI-alone counterfactual evaluation.
&
No assisted performance improvement in this setting.
&
\textbf{Cognitive change not measured.} The paper shows no performance benefit from recommendations, not that judges themselves remain cognitively unchanged.
&
Moderate methodological relevance. It demonstrates why the coupled decision process must be evaluated separately from the human or algorithm alone, but provides no evidence that the combination constitutes an extended cognitive system.
\\

\addlinespace

\textbf{Noy \& Zhang (2023)}\newline
\emph{Experimental Evidence on the Productivity Effects of Generative Artificial Intelligence \cite{noy_experimental_2023}}
&
Preregistered randomized online experiment with 444 college-educated professionals performing occupation-specific writing tasks; half receive ChatGPT access.
&
ChatGPT reduces completion time by 0.8 SD and increases output quality by 0.4 SD. Benefits are larger for lower-performing workers. The authors conclude that ChatGPT mainly substitutes for worker effort rather than complements worker skills.
&
Primarily the human worker under AI assistance.
&
Strong positive contemporaneous performance effect.
&
\textbf{Persistent improvement not measured.} The experiment establishes productivity augmentation during AI use, not acquisition of improved unaided writing ability.
&
Useful distinction between extension and internal improvement. The AI-supported configuration is more capable, but the results do not show that the human component itself has changed. This is compatible with a system-level interpretation, not proof of cognitive extension.
\\

\addlinespace

\textbf{Wang et al. (2025)}\newline
\emph{Exploring Creativity in Human--AI Co-Creation \cite{wang_exploring_2025}}
&
Human--AI Co-Creative Design Process (HAI-CDP) compared with a traditional design process; 42 designers, 21 novices and 21 experienced, performing an online concept-design task.
&
HAI-CDP substantially improves creative performance. Novices particularly increase idea generation; experienced designers gain more in quality and refinement. Experience remains important.
&
Human--AI co-creative interaction.
&
Positive assisted creative performance.
&
\textbf{Human improvement not established.} No subsequent AI-free creativity test is reported.
&
Relevant to distributed creative processing. Humans actively select, reject, and refine AI outputs rather than passively accepting them. This demonstrates interactive complementarity, but not that the combined system qualifies as an extended mind.
\\

\addlinespace

\textbf{Li et al. (2026)}\newline
\emph{Human-AI Collaboration Enhances the Performance of LLMs in Risk of Bias Assessment \cite{li_human-ai_2026}}
&
Two-stage evaluation. First, DeepSeek V3 and Qwen-plus assess 2,100 RoB samples independently. Second, junior human reviewers work under three Human--AI modes: evidence extraction (M1), reasoning support (M2), and disagreement-trigger intervention (M3). Accuracy, intervention rate, and intervention efficiency are measured.
&
AI-alone accuracy varies substantially; collaborative M2/M3 regimes improve accuracy, with reasoning support often producing the best accuracy and disagreement-triggering reducing human intervention. The authors conclude that LLMs cannot yet replace experts but can improve efficiency and reduce subjective bias in well-designed collaborative frameworks.
&
Human--AI interaction.
&
Positive collaborative-system performance.
&
\textbf{Human cognitive improvement not measured.} Reviewers are not tested later without AI.
&
Relevant to task-dependent integration and division of labour. Different collaboration architectures produce different joint-system properties. This supports studying integration structure, but not an Extended Mind conclusion.
\\

\addlinespace

\textbf{Aouad, Lykouris \& Zhong (2026)}\newline
\emph{Human-AI Productivity Paradoxes \cite{aouad_human-ai_2026}}
&
Theoretical economic/modeling paper in which human skill, effort, AI assistance, reliability, and AI literacy interact dynamically.
&
Under specified assumptions, increased AI assistance can reduce human effort; reduced effort can lower skill accumulation and, with unreliable AI, reduce long-run productivity. Heterogeneous AI literacy can generate skill polarization.
&
Modeled Human--AI dynamics.
&
Depending on parameters, short-run gains can coexist with long-run losses.
&
\textbf{No empirical human effect demonstrated.} Deskilling is a result of the theoretical model, conditional on its assumptions.
&
Strong conceptual relevance to temporal coupling. The human state changes as a consequence of prior AI interaction, which then alters future Human--AI performance. This is useful for treating cognition as a dynamic coupled process, but it is not an empirical test of Extended Mind.
\\

\addlinespace

\textbf{Wu et al. (2025)}\newline
\emph{Human-Generative AI Collaboration Enhances Task Performance but Undermines Human's Intrinsic Motivation \cite{wu_human-generative_2025}}
&
Four preregistered online experiments. Critically compares Collab$\rightarrow$Solo with Solo$\rightarrow$Solo, allowing performance during collaboration and subsequent unaided performance to be distinguished.
&
In all four studies, GenAI improves the initial assisted task. Study 1 finds positive spillover to idea creativity but not quantity. Studies 2--4 find no performance spillover on the subsequent solo task. Transitions from AI collaboration to solo work tend to reduce intrinsic motivation and increase boredom, although significance varies by study.
&
Both Human--AI interaction and the human after decoupling.
&
Positive while coupled; little general transfer once decoupled.
&
\textbf{Mixed but predominantly no improvement.} One limited positive creativity spillover; three studies show no general performance transfer. Psychological effects are partly negative.
&
One of the most relevant papers for an Extended Mind interpretation. It experimentally separates performance of the coupled configuration from performance of the human after decoupling. The augmentation appears largely configuration-dependent rather than internalized by the human.
\\

\addlinespace

\textbf{Dell'Acqua et al. (2026)}\newline
\emph{Navigating the Jagged Technological Frontier \cite{dellacqua_navigating_2026}}
&
Preregistered randomized laboratory-in-the-field experiment with highly skilled knowledge workers. GPT-only, GPT+overview, and control conditions; tasks deliberately selected inside and outside the AI capability frontier.
&
Inside the frontier, AI increases task completion by more than 12\%, improves quality by about 32\%, and reduces time substantially. Outside the frontier, AI assistance significantly reduces correctness.
&
Human performance under Human--AI interaction.
&
Positive or negative depending on task--AI fit.
&
\textbf{No persistent human change measured.} Current human performance can be either improved or degraded by AI assistance.
&
Important evidence against treating coupling as intrinsically beneficial. The performance of the coupled system depends on how the capabilities of its components fit the task. This is highly relevant to functional integration, although not evidence for constitutive extension.
\\

\addlinespace

\textbf{Riedl \& Weidmann (2026)}\newline
\emph{Quantifying Human-AI Synergy \cite{riedl_quantifying_2025}}
&
Two-stage Bayesian Item Response Theory applied to interactive Human--AI benchmark data ($n=667$). Separates individual ability from collaborative ability while controlling for task difficulty; also analyses Theory of Mind / perspective-taking.
&
Significant Human--AI synergy under the paper's definition of incremental value provided by AI to users. Collaborative ability is empirically distinct from individual problem-solving ability. Perspective-taking predicts superior Human--AI collaboration but not solo performance.
&
Human--AI interaction, with explicit modeling of human-alone ability.
&
Strong positive collaborative performance.
&
\textbf{No evidence that individual human ability improves over time.} There is no post-collaboration solo test establishing an increase in autonomous ability.
&
Probably the strongest empirical bridge to the Extended Mind question in this corpus. Collaborative ability differing from individual ability is consistent with treating $H+AI$ as a distinct functional configuration, but does not show that the configuration is a constitutive cognitive system.
\\

\addlinespace

\textbf{Du et al. (2026)}\newline
\emph{Systematic Literature Review of Human--AI Collaboration for Intelligent Construction \cite{du_systematic_2026}}
&
Systematic/bibliometric review. Searches Scopus, Web of Science, and Google Scholar; 301 initial documents and 191 retained after screening; bibliometric/co-occurrence analysis using VOSviewer and CiteSpace.
&
The literature is organized around robotics, productivity/safety, intelligent algorithms/modeling, and worker factors. The authors argue that Human--AI collaboration in construction requires consideration of physiological, psychological, and behavioural human factors, not just technology.
&
Human--AI interaction literature, rather than one empirical unit.
&
No single pooled performance result.
&
\textbf{No general improvement/deterioration conclusion is supported by the review itself.}
&
Limited/contextual relevance. It reinforces the importance of studying the whole sociotechnical interaction, but does not operationalize cognitive extension or establish a coupled cognitive unit.
\\

\addlinespace

\textbf{Str{\"o}mberg, Lei \& Wu (2026)}\newline
\emph{The Generative AI Learning Penalty \cite{stromberg_generative_2026}}
&
30-month longitudinal panel of more than 25,000 Chinese secondary-school students with staggered self-directed GenAI adoption. Difference-in-differences using Callaway--Sant'Anna regression adjustment. Measures homework productivity, monthly closed-book exams, and high-stakes entrance exams.
&
After about six months, homework scores rise 18\% and completion time falls from 64 to about 45 minutes. Simultaneously, monthly closed-book exam scores fall about 20\% of baseline; negative long-run entrance-exam effects also emerge more slowly.
&
Primarily the human learner over time, before and after AI adoption; AI-assisted work is contrasted with unaided assessments.
&
Positive assisted productivity but negative unaided learning.
&
\textbf{Deterioration in measured learning outcomes.} This is the strongest direct evidence in the corpus of a negative human effect, but the design is quasi-experimental rather than randomized and the conclusion is specific to this educational setting.
&
Highly relevant to the coupled-versus-internal distinction. Better performance of the AI-supported activity can coexist with weaker autonomous human capability. This is compatible with cognitive offloading/delegation, but does not establish that the Human--AI configuration is an extended mind.
\\

\addlinespace

\textbf{Vaccaro, Almaatouq \& Malone (2024)}\newline
\emph{When Combinations of Humans and AI Are Useful \cite{vaccaro_when_2024}}
&
Preregistered systematic review/meta-analysis of 106 experiments and 370 effect sizes, requiring human-alone, AI-alone, and Human--AI conditions. Uses three-level meta-analytic models and Hedges' $g$.
&
Human--AI systems outperform humans alone on average (human augmentation), but do not exhibit synergy on average relative to the better standalone agent. Decision tasks are associated with losses; creation tasks with gains. Robustness analyses reproduce the result.
&
Human--AI system, explicitly compared with human and AI components separately.
&
Positive relative to human alone; negative relative to the best standalone component on average.
&
\textbf{Human cognitive improvement not measured.} ``Human augmentation'' in this paper is a performance comparison, not retained human learning.
&
Very strong methodological relevance. It explicitly evaluates $H$, $AI$, and $H+AI$, making the unit-of-analysis issue visible. Its results also show that mere coupling does not guarantee emergence or synergy. It does not claim Extended Mind.
\\

\addlinespace

\textbf{Srivastava et al. (2023)}\newline
\emph{Beyond the Imitation Game: Quantifying and Extrapolating the Capabilities of Language Models / BIG-bench \cite{srivastava_beyond_2023}}
&
204 tasks contributed by 450 authors across 132 institutions. Evaluates GPT, Google dense/sparse models, and PaLM across model sizes. Expert humans perform tasks to provide a strong baseline. Zero/few-shot performance, calibration, scaling, and task-specific behaviour are analysed.
&
Model performance and calibration generally improve with scale but remain below expert-rater aggregate performance in the models tested. Apparent ``breakthrough'' capabilities can depend strongly on evaluation metric and task formulation; exact-match metrics can make gradual underlying changes appear abrupt.
&
The AI model. Humans are baseline comparators, not interaction partners.
&
Model performance generally improves with scale; this is not a Human--AI result.
&
\textbf{Not measured at all.} The paper provides no evidence about whether AI interaction improves, worsens, or leaves a human unchanged.
&
Methodological rather than empirical relevance to Extended Mind. BIG-bench exemplifies model-centred evaluation and cannot capture capacities that might exist only in an interactive Human--AI configuration. It can motivate, but not validate, a change in unit of analysis.
\\

\end{longtable}
\end{landscape}

\clearpage

\end{document}